\documentclass[useAMS,usenatbib]{mn2e}
\usepackage{gensymb}
\usepackage{amsmath}
\usepackage{amssymb}
\usepackage{mhchem}
\usepackage{color}
\usepackage{graphicx}
\usepackage{natbib}
\usepackage{float}
\usepackage{subfig}
\usepackage{aas_macros}
\newcommand{\kms}{~km\,s$^{-1}$}
\newcommand{\ms}{~m\,s$^{-1}$}
\newcommand{\logg}{$\log\,g$} 
\newcommand{\teff}{T$_{\rm{eff}}$}

\title[The GALAH Survey Motivation]{The GALAH Survey: Scientific Motivation}
\author[De Silva, Freeman, Bland-Hawthorn et al.]
{G.M. De Silva$^{1,3}$\thanks{Email:  \texttt{gayandhi.desilva@aao.gov.au}}, 
K.C. Freeman$^2$\thanks{Email:  \texttt{kenneth.freeman@anu.edu.au}}, 
J. Bland-Hawthorn$^3$\thanks{Email:  \texttt{jbh@physics.usyd.edu.au}}, 
S. Martell$^4$, E. Wylie de Boer$^2$, 
\newauthor{M. Asplund$^2$, S. Keller$^2$, S. Sharma$^3$, D.B. Zucker$^{5,6}$, T. Zwitter$^7$,
B. Anguiano$^{5,6}$,}
\newauthor{C. Bacigalupo$^{5,6}$, D. Bayliss$^{2}$, M.A. Beavis$^8$, M. Bergemann$^9$, 
S. Campbell$^{10}$, R. Cannon$^1$,}
\newauthor{D. Carollo$^{19,20}$, L. Casagrande$^2$, A.R. Casey$^{11}$, G. Da Costa$^2$, V. D'Orazi$^{12,5}$, A. Dotter$^2$,}
\newauthor{L. Duong$^2$, A. Heger$^{10,17,18}$, M.J. Ireland$^2$, P.R. Kafle$^{13}$, J. Kos$^7$, J. Lattanzio$^{10}$,} 
\newauthor{G.F. Lewis$^3$, J. Lin$^2$, K. Lind$^{14}$, U. Munari$^{15}$, D.M. Nataf$^2$, S. O'Toole$^1$, Q.A.Parker$^{5,6}$,} 
\newauthor{W. Reid$^{5,6}$, K.J. Schlesinger$^2$, A. Sheinis$^1$, J.D. Simpson$^{5,6}$, D. Stello$^3$, Y-S. Ting$^{16}$,} 
\newauthor{G. Traven$^7$, F. Watson$^1$, R. Wittenmyer$^4$, D. Yong$^2$, M. \v Zerjal$^7$
}
\\\\
(Affiliations listed after the references)}

\begin{document}

\date{Accepted TBC. Received TBC; in original form TBC}

\pagerange{\pageref{firstpage}--\pageref{lastpage}} \pubyear{2015}

\label{firstpage}

\maketitle

\begin{abstract}
The GALAH survey is a large high-resolution spectroscopic survey using the newly commissioned HERMES spectrograph on the Anglo-Australian Telescope. The HERMES spectrograph provides high-resolution (R $\sim$28,000) spectra in four passbands for 392 stars simultaneously over a 2 degree field of view. The goal of the survey is to unravel the formation and evolutionary history of the Milky Way, using fossil remnants of ancient star formation events which have been disrupted and are now dispersed throughout the Galaxy. Chemical tagging seeks to identify such dispersed remnants solely from their common and unique chemical signatures; these groups are unidentifiable from their spatial, photometric or kinematic properties. To carry out chemical tagging, the GALAH survey will acquire spectra for a million stars down to V $\sim$14. The HERMES spectra of FGK stars contain absorption lines from 29 elements including light proton-capture elements, $\alpha$-elements, odd-Z elements, iron-peak elements and n-capture elements from the light and heavy s-process and the r-process. This paper describes the motivation and planned execution of the GALAH survey, and presents some results on the first-light performance of HERMES.
\end{abstract}

\section{Introduction }
\label{sec:intro}

About half of the stars in the Milky Way are thought to have formed before an age corresponding to redshift z $\sim$1 \citep{lineweaver}. We see ancient stars around us in the old thin disc, the thick disc, the stellar halo, the inner bulge, in globular clusters and in satellite dwarf galaxies. We are coming into a new era of Galactic investigation, in which one can study the fossil remnants from the earliest phases of the Galaxy's formation throughout its major luminous components. The earliest galaxy formation scenario by \cite{els} suggested the dissipative collapse of a large isolated protogalactic gas cloud, which eventually settled into a disc. The idea of galaxy formation through hierarchical aggregation of smaller elements from the early Universe later gained support from observational \cite[e.g.][]{searlezinn} and theoretical \cite[e.g.][]{peebles71,press_schechter} perspectives. Today we acknowledge that both processes of dissipative collapse and satellite accretion play a key role in the formation and evolution of galaxies. The influence of infall of small evolved satellite systems is not confined to the stellar halo: it may also affect the build-up of the disc and bulge. Such infall is prominent in Cold Dark Matter (CDM) simulations of galaxy formation \citep[see for example][]{abadi}, and we need to consider the detection of the debris of such systems in the context of the archaeology of the Galaxy's major components.

Identifying the debris of smaller building blocks of the Galaxy offers the possibility to reconstruct at least some properties of the proto-galaxy and so to improve our basic understanding of the galaxy formation process. This idea was present in early studies of moving stellar groups \citep{eggen74}, and is now becoming a major field of research in theoretical and observational astronomy. The goal of these studies in Galactic archaeology is to unravel the early Galactic history by using the stellar relics of ancient in-situ star formation and accretion events, primarily focused on the building-up of the disc and bulge, which contain nearly all of the stellar mass of the Milky Way. 

That said, unraveling the history of disc formation is likely to be challenging as much of the dynamical information such as the integrals of motion is lost due to heating. We need to examine the detailed chemical abundance patterns in the disc components to reconstruct substructure of the protogalactic disc. Pioneering studies on the chemodynamical evolution of the Galactic disc by \cite{ed93} followed by many other such works \citep[e.g.][]{reddy2003,bensby2014study}, show how trends in various chemical elements can be used to resolve disc structure and obtain information on the formation and evolution of the Galactic disc, e.g. the abundances of thick disc stars relative to the thin disc. The effort to detect relics of ancient star formation and the progenitors of accretion events will require gathering kinematic and chemical composition information for vast numbers of Galactic field stars.

\cite{fbh02} proposed the method of chemical tagging in order to reconstruct some of the ancient star-forming aggregates in the disk. The goal of the Galactic Archaeology with HERMES (GALAH)\footnote{http://www.mso.anu.edu.au/galah/home.html} survey is to acquire high-resolution spectra of about a million stars for chemical tagging, in order to investigate the assembly history of the Galaxy. This paper describes the scientific motivation for the GALAH survey, starting with some general points about the main components of the Galaxy in Section \ref{sec:galactic_components}. In Section \ref{sec:chemicaltagging}, we discuss some specifics of chemical tagging. Section \ref{sec:hermes} presents a summary of the HERMES spectrograph, which is the workhorse instrument of the GALAH survey. Section \ref{sec:galah} provides an overview of the survey operations and in Section~\ref{sec:scope} we discuss survey scope and scientific opportunities.

\section{Galactic Components}
\label{sec:galactic_components}

In this section we present a brief overview of the major Galactic components in order to provide a context for the later discussion of the GALAH survey scope.

\subsection{The Halo and its Satellites}
The metal-poor stellar halo is a minor component in terms of the relative stellar mass content of the Galaxy. It hosts some of the oldest Galactic stellar populations and is dynamically and chemically very different from the rest of the (stellar) Galaxy \citep{amina08}. It is spatially very extended and slowly rotating. The dynamical timescales are long and dynamical interactions less severe than in the disc. Halo stars can therefore preserve their integrals of motion more readily and their dynamics are more likely to contain useful information about the assembly of the halo. For discussions on the state of the Galactic halo based on Sloan Digital Sky Survey (SDSS) see e.g. \cite{carollo, carollo10, beers, fermani, kafle, schonrich}.

We suspect that a large fraction of the halo stars may be remnants of early satellite galaxies which 
experienced independent chemical evolution before being accreted on to the Galaxy \citep{kirby11}. 
The discovery of many dwarf satellites and stellar streams including the Sagittarius dwarf spheroidal \citep{ibata94}, is evidence that such processes are still underway \citep{zucker06}. The stars in dwarf spheroidal galaxies generally have lower $\alpha$-element abundances compared to the halo stars, except at very low metallicities ([Fe/H] $\lesssim -1.5$), where the $\alpha$-element abundances overlap with those of halo stars \citep{tolstoy09}. Identifying any disrupted primordial counter-parts of the dwarf satellites currently orbiting the Galaxy, will serve as probes into the conditions of halo assembly as well as star formation in the early Universe. 

Globular clusters represent some of the oldest stellar populations in the Galaxy with characteristic photometric 
and light element abundance peculiarities including the O-Na anti-correlation \citep{gratton2012}. Stars with such particular light-element abundance patterns are observed mostly within globular clusters, but not generally in the field \citep[e.g.][]{gratton2000} or in Local Group dwarf galaxies  \citep[e.g.][]{tolstoy09}. The small fraction of such stars in the field \citep[e.g.][]{carretta2010,martell11,wdb12} are interpreted as evidence of cluster mass loss and dissolution. Globular cluster formation and origin is a subject of active research and the detection of extra-tidal tails and field stars that can be chemically tagged back to globular clusters will provide significant new insights.

\subsection{Thin and Thick Discs}

The Galactic thin disc shows a mean radial metallicity gradient of about $-$0.07 dex
kpc$^{-1}$ \citep[e.g.][]{cheng12}. Near the Sun, the thin disc stars cover a wide range in age, with the oldest stars having ages up to about 10 Gyr. Their age-metallicity relation shows that stars of almost all ages up to about 10 Gyr cover a broad metallicity range (from about $-0.5$ to +0.5 dex), and only a weak trend of increasing metallicity with decreasing age is discernible. The current belief is that the most metal-rich thin disc stars near the Sun did not form near the Sun but rather formed in the more metal-rich inner Galaxy and migrated radially out to the solar neighbourhood \citep[e.g.][]{grenon,haywood08}. Radial migration is a theoretical prediction that has generated a great deal of interest recently. It is believed to be driven by the torques of the Galactic bar and transient spiral arms, which can move stars radially inwards and outwards from one near-circular orbit to another \citep{sellwood_binney}, without significantly heating the disc. We do not know how important radial mixing has been in determining the current state of the Galactic disc. The GALAH survey data will be able to answer this question (see Section \ref{sec:radialmigration}).

In addition to their defining thin discs, almost all spirals appear to have a thick disc component \citep{yoachim06}. The Galactic thick disc was originally identified through vertical star counts \citep{yoshii,gilmore83}, and has since been shown to be generally older, more metal-poor and more $\alpha$-enhanced than the thin disc, though there are also old metal-rich stars with thick-disc-like orbits (i.e. large scale heights and slower
rotational velocity). Thus, it is an open question whether the stellar disc can form a thick component with time because of internal (e.g., scattering, radial mixing) or external (accretion events, mergers) mechanisms or is a continuous distribution \citep{bovy12}.

For large spirals like the Milky Way, the thick disc mass is typically about 10\% of the thin disc in the solar neighbourhood, and its vertical scale height is about 1000 pc; the scale height of the thin disc is typically about 300 pc \citep{casagrande11}. Stars of the thin and thick discs have different motions and different density distributions. For a recent assessment of disc kinematic parameters see \cite{sharma14}. 

Near the Sun, although the thick disc is on average more metal poor, the metallicity ranges of the thick and thin disc stars overlap. The thin disc stars have [Fe/H] in the range $-$0.7 to +0.5, while most of the thick disc stars have [Fe/H] between about $-$1.0 and $-$0.3, with tails extending to $-$2.0 and $-$0.1 \citep{schlesinger}. In the [Fe/H] interval over which the two disc components overlap, the thick disc stars are distinctly more $\alpha$-enhanced \citep[see e.g,][]{bensby2014study}. Many fundamental questions remain unanswered, such as the metal-poor limit of the thin disk, and the metal-rich limit of the thick disk. What is the radial and vertical extent of the thick disc and do they display any metallicity gradients? With a simple selection function and large sample size, the GALAH dataset will play a crucial role in answering these basic questions (see Section \ref{sec:thinthick}).

\subsection{Bulge}

The bulge of our Galaxy is archaeologically interesting. Small boxy bulges like the bulge of the
Milky Way are now not regarded as predominantly being the products of mergers. The Milky Way bulge is believed to have formed several Gyr ago via bar-forming and bar-buckling instabilities of the early disc \citep[e.g.][]{athanassoula}. The disc forms an elongated bar structure at its center, which then buckles vertically and settles into the long-lived boxy shape. These instabilities of the disc redistribute the disc stars into the bulge. This picture is supported by comparisons of radial velocity observations to N-body model predictions \citep[e.g.][]{shen}. The different components of the early inner disc (thick disc, old thin disc, younger thin disc) end up trapped dynamically within the bulge structure \citep{ness13a}. Their distribution within the bulge
depends on their initial phase space distribution before the instability. We can see the relics of
these Galactic components now in the metallicity distribution function of the bulge stars \citep[see Fig 1 in][]{ness13a}. These metallicity components are seen throughout the bulge, but their relative weights change with position in the bulge. In this way, we see a fossil image of the early inner Galaxy, mapped into the bulge.

Although the bulge giants are mostly too faint (V $>$ 14) for the GALAH survey, dedicated bulge programs using the HERMES instrument have the potential to contribute very significantly to understanding the chemical evolution of the bulge and inner disk.

\section{Chemical Tagging}
\label{sec:chemicaltagging}

The idea of chemical tagging is to use the detailed chemical abundances of individual stars to tag or associate 
them to common ancient star-forming aggregates (including disrupted satellites) whose stars have similar abundance patterns \citep{fbh02}. After their originating clusters or host galaxies disperse, these stars may lose much of their dynamical information, such as the integrals of motion, because of heating and radial migration. The stellar abundance pattern over many elements likely reflects the unique chemical state of the gas from which the aggregate formed, analogous to a star's DNA profile.

\subsection{Open clusters}
A vital condition for the chemical tagging process is that star clusters be chemically homogeneous, and their abundance distributions be sufficiently different from cluster to cluster \citep{desilva2007, desilva2009}. Almost all surviving open star clusters are chemically homogeneous at the level of the measurement uncertainties ($\sim$0.05 dex) for heavier elements and often also for the lighter elements in stars whose surface abundances are not affected by stellar evolutionary effects such as dredge-up or deep mixing (see \cite{friel2014} for a recent example). Careful differential analyses of high-resolution data with high signal-to-noise, which removes much of the line-to-line systematics, show abundance scatter at the level of $\sim$0.02 dex, where the source of chemical inhomogeneity in open clusters could include effects of planet formation \citep{ramirez} and/or atomic diffusion \citep{onehag}. Such levels of scatter may be the limit at which we can chemically resolve individual star formation events. 

Theoretically, a high level of chemical homogeneity is expected even within regions of low star formation efficiency that do not produce bound open clusters \citep{krumholz2014}. Dynamically, open clusters that remain bound over a Gyr may be anomalous disc objects. Star clusters that disperse within the molecular clouds may have been the major contributors to the building of the Galactic disc \citep{lada2003}. Enhanced [Na/Fe] in old open cluster stars have been reported in the literature \citep[e.g.][]{jacobson2007study, schuler2009study} whereas solar level Na abundances are reported in field star studies. This result questioned whether old open clusters are chemically different to the star formation events that contributed to the Galactic disc. However, this discrepancy in Na abundances is likely to be a result of analysis method, where departures from local thermodynamic equilibrium were not taken into account in earlier open cluster studies \citep{maclean}. 

\subsection{Moving groups}

The Galactic disc near the Sun shows some kinematic overdensities or substructure. These are sets of stars that have some degree of common motion, and they are broadly referred to as stellar moving groups. The stars of the moving groups are all around us: they are seen as concentrations in velocity but not in position. Some of these groups are the debris of disrupted old star clusters in the disc. Examples include the HR 1614, Wolf 630 and Argus moving groups \citep{desilva2007, bubar, desilva13}. These groups are now dispersed into extended regions of the Galaxy, and their stars are chemically homogeneous to within 0.05 dex and have common ages. The HR 1614 and Wolf 630 groups are about 2 Gyr old but still retain some kinematic identity. We can expect to find many older disrupted groups which have lost all their kinematic identity, distinguished by the fact that they preserve their chemical identity. These are the relics that trace the star formation history of the Galaxy, and are of great interest for Galactic archaeology.

Other moving groups, such as the Hercules group \citep{bensby07}, represent dynamical resonances
associated with the pattern speed of the Galactic bar or spiral structure. The Hercules group members are a chemically typical sample of the nearby disc, with a wide range of chemical abundances. These resonance groups are dynamically interesting but are of limited interest for Galactic archaeology. Further chemical information shows that a co-moving group of stars can have multiple origins, such as the Hyades Supercluster. In this case some of the dispersed stars share identical abundances to that of the Hyades open cluster, while the majority of the co-moving stars present the field star abundance distributions \citep{desilva11, pompeia,tabernero}, showing that the Hyades Supercluster has a combined dissipative and dynamical origin.

\subsection{In-falling satellites}

Some moving groups may be the debris of in-falling dwarf galaxies that were tidally disrupted in the process of being accreted by the Milky Way. The disrupting Sagittarius dwarf galaxy is a familiar example \citep{ibata94}. The Kapteyn group is another example, as a possible remnant of the tidal debris of the unusual globular cluster Omega Centauri during a merger event that deposited Omega Cen into the Milky Way  \citep{wdb10, majewski}. One of the GALAH survey goals is to identify such debris using chemical techniques, even if the debris has lost its kinematic identity, because this provides a way to make a direct estimate of the accretion history of the Milky Way.

In summary, although the disc does show some surviving kinematic substructure in the form of moving stellar groups, we can expect that much of the dynamical information was lost in the subsequent heating and radial mixing by spiral arms and interactions with giant molecular clouds. Groups like the HR 1614 group are rare examples of dispersed clusters which cannot be identified spatially but are still identifiable both chemically and kinematically. Most older dispersed aggregates would now not be recognisable dynamically, and chemical techniques provide the only way to identify their debris.

\subsection{C-space}

The process of chemical tagging operates within a multidimensional space (C-space) whose axes are the measured relative abundances of the stars being studied. The C-space co-ordinates are [X$_{1}$/H], [X$_{2}$/H], ... [X$_{n}$/H]. From HERMES spectra of GALAH targets, abundances for up to 29 elements per star will be measurable. Not all of these elements vary independently from star to star; many vary together in near-lockstep. The number of independent dimensions of this chemical space appears to be between 7 to 9, possibly depending on the dominant nucleosynthetic processes in a given Galactic population \citep{ting12}. 

Stars that originate from chemically homogeneous aggregates like dispersed clusters will lie in tight groups in C-space. Stars which were accreted from dwarf galaxies will lie along chemical sequences that reflect the unique history of star formation and chemical enrichment in those galaxies. We expect these sequences will be distinct in C-space from the stars of the Galactic thin and thick discs. We expand the discussion on the dimensionality of C-space and required sample size for chemical tagging in Section~\ref{sec:scope} in relation to the scope of the GALAH survey.

With this chemical tagging approach, we expect that it will be possible to reconstruct old dispersed stellar aggregates in the Galactic disc, derive their ages and their kinematic and metallicity distributions, and so build up a picture of the star formation history and the dynamical and chemical evolution of the Galactic disc since the disc began to assemble. This kind of chemical tagging experiment needs a high-resolution spectroscopic survey of about a million stars, homogeneously observed and analysed - this is the GALAH survey. It is the prime science driver for the HERMES instrument.

\section{\bf The HERMES Spectrograph}
\label{sec:hermes}

A million star survey as discussed in this paper is only possible with a spectroscopic facility combining 
high resolution and high multiplex over a large field of view and sufficient spectral coverage in a single exposure. This facility is the High Efficiency and Resolution Multi-Element Spectrograph (HERMES) recently commissioned on the AAT. It is currently unrivalled as the only optical spectrograph with this combination of capabilities. In this section we briefly describe the 2dF+HERMES design and performance. For full details of HERMES functionality and commissioning results see \cite{sheinis}.

HERMES receives light from the existing 2-degree Field (2dF) fibre positioning system, mounted at the prime focus of the AAT, which collects light into 400 fibres over a circular field 2 degrees in diameter \citep{2df}. Eight of these fibres are fiducial fibres used to align and guide the observing field on user provided catalogue stars. The remaining 392 fibres are available for science use, of which at least 25 are recommended for sky background observations. The fibre diameter is 2.1 arcsec and the robotic positioner has a fibre positioning accuracy of 0.3 arcsec.

The two-prism atmospheric dispersion corrector (ADC) at the prime focus of the AAT automatically counteracts the colour spread in each star image due to atmospheric dispersion. There are also positional errors due to differential atmospheric refraction across the 2dF field during long exposures: these are minimised by the customised 2dF fibre configuration software and by observing close to the meridian. Third, due to its original design requiring a flat field and telecentric input beams, there is systematic chromatic variation in distortion (CVD) across the 2dF field. The effect is negligible at the center of the field, where the light from different wavelengths coincides. At larger radii on the 2dF plate, the light is spread radially up to a maximum of about 0.6 arcsec between the centroids at 5250\AA\ and 8500\AA\ \citep{cannon08}. Towards the edges of the field the radial shifts decline but the images become more diffuse. Thus there are three colour-dependent effects that all contribute to the fibre positioning accuracy and hence to the effective throughput of 2dF plus HERMES. The first is largely corrected by the ADC, the second is minimised by the configuration software and careful planning of the observations, but the third is a feature of the system which constrains the performance across
wide wavelength ranges.

The fibre cable of almost 50 meters in length runs from the telescope prime focus into a stable, temperature-controlled room inside the AAT West Coude laboratory where the HERMES spectrograph is located. The 2dF positioning system contains two observation plates, which allows the fibre configuration to be updated on one plate while the other plate is collecting data. Correspondingly, the HERMES slit assembly holds two interchangeable slit units, which provide an accurate and stable interface for the two fibre feeds coming from 2dF. 

Post slit, HERMES has a single off-axis collimator and two off-axis corrector lenses, four VPH (Volume Phase Holographic) gratings and cameras and three dichroic beam splitters. Each camera feeds one 4096$\times$4112 15$\mu$m pixel CCD from the E2V CCD231-84 family. The Blue and Green channel detectors are both 16 micron standard silicon devices. The Red channel detector is a 40 micron, deep depletion device and the "IR" channel detector is a 100 micron bulk silicon device. Both Red and IR CCDs have fringe suppression coatings. Each detector is independently controlled and permits readout from one, two or four readout amplifiers, at various clock rates with windowing and binning options. 

HERMES provides fixed format spectra in four non-contiguous wavelength bands covering a total of about 1000\AA\ between 4713\AA\ and 7887\AA. HERMES is designed to operate at two resolution modes, with a resolving power of R $\sim$ 28,000 being the nominal resolution mode and a higher resolving power of R $\sim$ 42,000 achieved by manually inserting a mask on kinematic mounts on the slit assembly, at the cost of 50\% light loss. The standard HERMES setup for GALAH observations uses the nominal resolution mode with two amplifiers in fast readout (4e readout noise) for all CCDs. Further details on the HERMES design and performance see \cite{barden2010, heijmans2012, sheinis}.
 
\begin{figure*}
\centering
\includegraphics[angle=270,width=8cm]{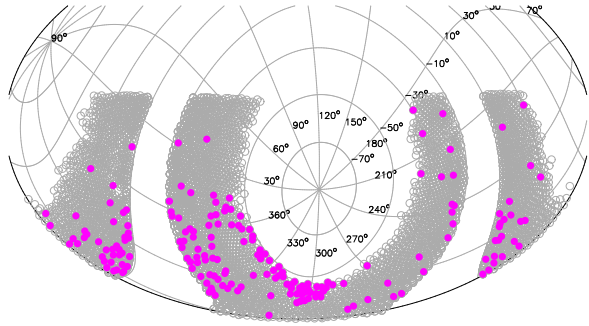}
\caption{The distribution of the GALAH observing fields. There are 4303 field centers shown as grey open circles, of which 3300 need to be observed to complete the survey. The magenta circles show the observations completed as at November 2014.}
\label{fig:tiles}
\end{figure*}

\section{Survey Observations}
\label{sec:galah}

\subsection{Target selection}

The GALAH survey plans to observe $\sim$1,000,000 stars, using AAT+HERMES in the nominal resolution mode. Our priority is to keep the GALAH selection criteria as simple as possible, with a baseline selection being a magnitude-limited sample corresponding to 12$<$V$<$14 (with some excursions brighter and fainter for special targets) with $\vert b\vert >$ 10 degrees and declination between $-$80$\degree<\delta<$ +10\degree. We use the Two Micron All Sky Survey (2MASS)\footnote{http://www.ipac.caltech.edu/2mass/} JHK photometry \citep{2mass} to estimate V(J,K) to define our target selection criteria, where
$$
 V(J,K) =  K + 2(J-K+0.14)+0.382e^{((J-K-0.2)/0.50)}. 
$$
 The AAVSO Photometric All-Sky Survey (APASS\footnote{http://www.aavso.org/apass}, \cite{henden,henden14,munari14}) optical photometry as available, is used to estimate the stellar magnitudes in the HERMES bands to check throughput estimates. These, together with 2MASS photometry act as photometric priors for the subsequent spectroscopic analysis, e.g. using the Infrared Flux Method \citep{casagrande10}. 

The above mentioned selection criteria is used to choose fields with target densities $>$ 400 stars per 2dF field to allow optimal allocation of fibres. Using an adaptive tiling strategy\footnote{Tiling refers to the layout of 2-degree fields on sky. Details of the tiling strategy are given in Sharma et al. (in prep).} , the input catalogue is divided into 4303 field centers which in turn yield 6546 independent 'configurations' containing 400 unique stars. Of these, 3300 configurations need to be observed in order to reach the target number of stars (see Figure \ref{fig:tiles}). 

\begin{figure*}
\centering
\includegraphics{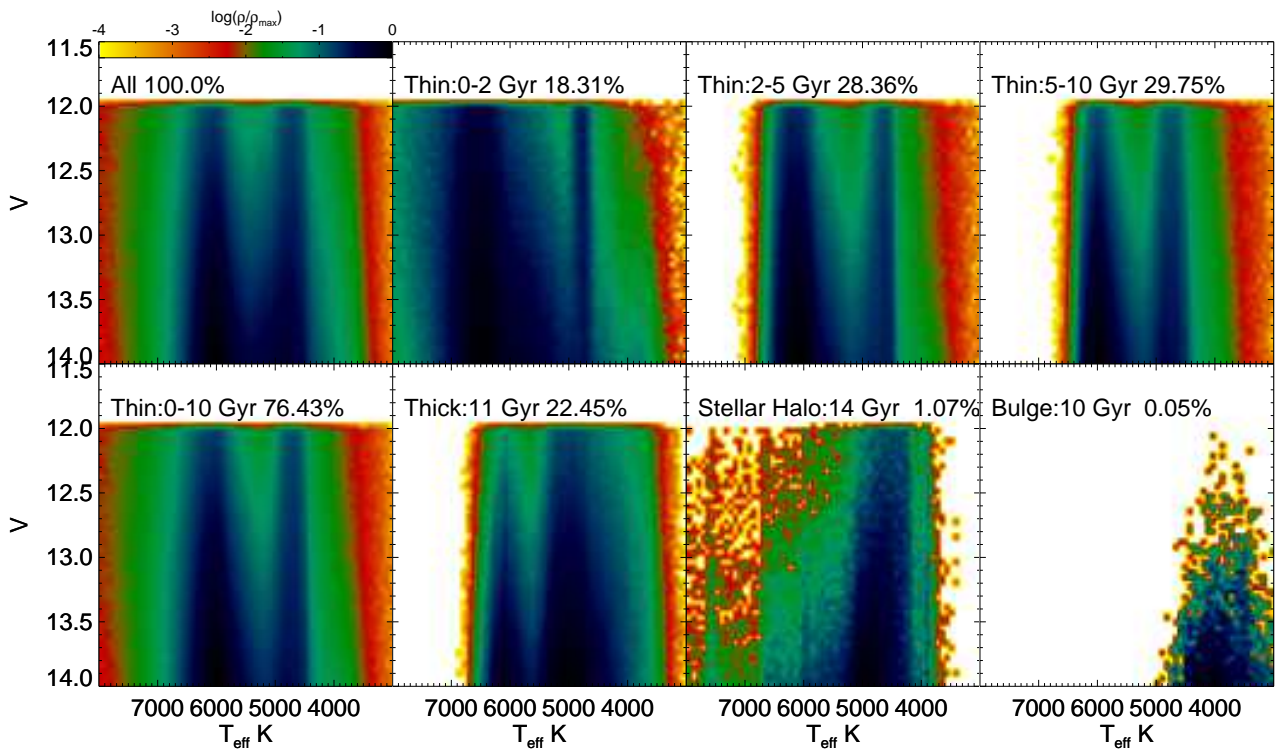}
\caption{The effective temperature vs. magnitude diagram of different galactic populations from a synthetic GALAH survey, 
based on Galaxia \citep{galaxia}.}
\label{fig:distribution}
\end{figure*}

The stars in such a magnitude-limited survey will have a double-peaked temperature distribution, with one peak dominated by stars near the main sequence turnoff and the other by clump giants (see Figure \ref{fig:distribution}). Table \ref{tab:sample} provides an estimate of the expected contribution from the giants and dwarfs from each of the main Galactic components, based on Galaxia models \citep{galaxia}. The old disc dwarfs are seen out to distances of about 1 kpc, the clump giants to about 5 kpc, and the brightest halo giants to about 15 kpc.

The {\sc ObsManager} tool developed within the GALAH collaboration interfaces with the GALAH input catalogue to provide observers with 2dF configuration files appropriate to a particular date and time. For each field center in our observing plan, {\sc ObsManager} will select as potential target stars only those with no close neighbours. For this selection we used the 2MASS {\it prox} parameter, which is the distance between source and nearest neighbour up to the catalogue magnitude limit, where we set a distance limit of {\it prox} $>$ 6 arcsec for all stars and distance greater than $130-(10\times V)$ arcsec for brighter stars. We use the Tycho and Bright Star Catalogs, with larger exclusion zones for brighter stars. {\sc ObsManager} also ensures a distance of at least 30 degrees from the Moon, and avoids fields containing Solar System planets to prevent the contamination across large areas of the observing field. Since the astrometry in our catalog is taken from 2MASS, {\sc ObsManager} includes the positions of targets along with their proper motions from the UCAC4 \citep{ucac4} catalogue when writing out 2dF configuration files. 

Upon completion of observations (see section \ref{section:dataflow}), {\sc ObsManager} updates the input catalogue to prevent reallocation of observed targets in future field configurations. Targets marked for repeat observations, individual targets (such as calibration standard stars) and high priority targets (such as synergistic fields, see section \ref{sec:synergy}) are also handled within {\sc ObsManager}, to optimally integrate with regular GALAH field configurations. A detailed description of the {\sc ObsManager} software is presented in Sharma et al. (in prep).

\begin{table}
\centering
\caption{Percentage of contribution to the GALAH sample from Galactic components.}
\label{tab:sample}
\begin{tabular}{ccc}
\hline
\hline
Component & Dwarf & Giant  \\
\hline
Thin disc & 57.87 & 18.56 \\ 
Thick disc & 4.63 & 17.82 \\
Halo & 0.05 & 1.02 \\
Bulge & 0.00 & 0.05 \\
\hline
\hline
\end{tabular}
\end{table}

\subsection{Baseline data flow}
\label{section:dataflow}

The baseline GALAH observing strategy is to observe each configuration for three consecutive twenty-minute exposures within an hour of transit. In practice, the total exposure time is dependent on the seeing, with an additional 20 minutes for seeing larger than 2 arcsec and an additional 60 minutes for seeing larger than 2.5 arcsec necessary to meet our S/N requirement of 100 per resolution element. When seeing conditions are worse than 3.0 arcsec, observation of regular GALAH fields are halted, although bright calibration observations, such as Gaia benchmarks stars \citep{jofre14} can be carried out if clear skies prevail.

Our observing strategy also includes repeat observations of 5\% of GALAH targets. Repeat observations allow us to monitor the quality of the collected data and the reliability of the data reduction, to determine the internal errors in our data-analysis pipeline, and to identify variable sources in our sample. 

GALAH standard calibration observations consist of one UV-Quartz flat-field lamp exposure and one ThXe arc lamp exposure per 2dF configuration, taken immediately before or after the science observations to ensure a similar telescope position as the target exposure. A set of 15 bias frames are taken every afternoon and dark frames are taken intermittently, typically during bad weather. Twilight sky flats and dome flats improve the flat-fielding accuracy and are taken at all available opportunities.

The data reduction and analysis is managed by the GALAH Analysis Pipeline (GAP) developed by the GALAH collaboration. The main stages of GAP consists of a series of modules that perform the required data reduction, quality control, chemical abundance determination and data product delivery. 

As new data becomes available, GAP makes use of the 2dfdr\footnote{www.aao.gov.au/science/software/2dfdr} automatic data reduction pipeline dedicated to reducing multi-fibre spectroscopy data, as well as an IRAF\footnote{IRAF is distributed by the National Optical Astronomy Observatories, which are operated by the Association of Universities for Research in Astronomy, Inc., under cooperative agreement with the National Science Foundation.}-based reduction tool developed within the GALAH collaboration with pre-existing IRAF routines, adapted to reduce HERMES multi-fibre spectra. The dual reduction outputs are maintained from this point onwards to check the reduction quality and to aid in the on-going improvements to the reduction procedures.

Following data reduction, quality checks including the resolution and signal-to-noise ratio (SNR) is determined for each individual spectrum. Rapidly rotating stars and unusual stars like spectroscopic binaries and chromospherically active stars are identified using the local linear embedding techniques such as by \cite{matijevic2012} and set aside. Fields with a sufficient number of stars that meet the data quality requirements are considered to be 'completed' and removed from the input catalogue by {\sc ObsManager}. 

For all stars that pass the quality checks, the spectra from the four wavelength bands are processed through the abundance analysis module. In summary, this module carries out the radial velocity and continuum normalisation of the spectra, determines the stellar parameters based on spectroscopic techniques and available photometric priors, and finally calculates individual elemental abundances via spectral synthesis. The choice of atomic line data and model atmospheres are user provided inputs. Detailed information on the analysis pipeline process is given in Wylie de Boer et al. (in prep).

\subsection{\bf GALAH spectra}

The HERMES throughput requirement for GALAH science is to obtain a minimum SNR of 100 per resolution element (where the resolution element is approximately 4$\times$4 pixels) for a 14$^{th}$ magnitude star in the corresponding channel central wavelength within one hour exposure time. Note the nearest Johnson-Cousins filters for the HERMES channels are B,V,R and I. Based on GALAH data, Table~\ref{tab:efficiency} presents the limiting magnitude to achieve the required signal-to-noise in one hour exposure under seeing conditions of 1.5 arcsec for each HERMES channel. With the instrument achieving the required throughput, we expect the bulk of the survey data (12$<$V$<$14) to meet the signal-to-noise requirements. Note that for a given star, the signal-to-noise across the four channels will depend on the stellar type; a hotter star will have more signal in the bluer wavelength channels compared to a cooler star of the same V magnitude.
 
\begin{table}
\centering
\caption{Magnitude limit to obtain signal-to-noise of 100 per resolution element in 1 hour exposure}
\label{tab:efficiency}
\begin{tabular}{|c|c|c|c|c|c|c|c|c|c|}
\hline
Channel  & $\lambda_{\rm{min}}$ (\AA) & $\lambda_{\rm{max}}$ (\AA) & Magnitude \\
\hline
Blue & 4713 & 4903  & B = 14.2  \\ 
Green & 5648 & 5873 & V = 13.8  \\
Red & 6478 & 6737  & R = 14.0 \\
IR & 7585 & 7887 & I = 13.8 \\
\hline
\end{tabular}
\end{table}

\begin{figure}
\centering
\includegraphics[angle=270,width=8cm]{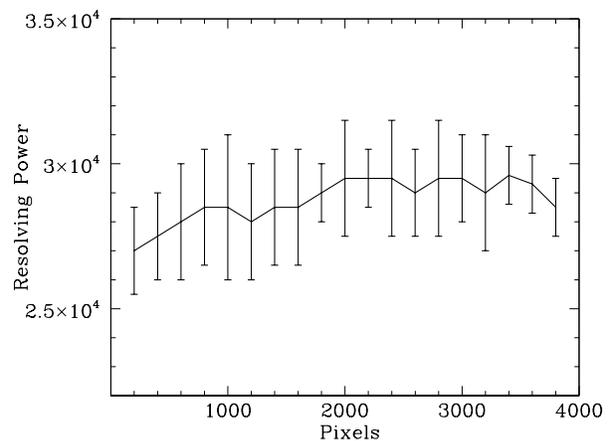}
\caption{Typical resolving power measured from the central fibre against spectral pixel position measured from a typical GALAH observational field.  The error bars represent the range in resolving power between the four channels and between fibres.}
\label{fig:resolution}
\end{figure}

The achieved spectral resolution in the GALAH data for each HERMES channel is R $\sim$ 28,000. The resolution was measured by fitting gaussian profiles to ThXe calibration arc lamp exposures. The typical values measured from the central fibre are presented in Figure \ref{fig:resolution}. However the resolving power varies between fibres and between the four channels. The combined range of variations in resolving power are represented by the error bars.

\begin{figure}
\centering
\includegraphics[angle=270,width=8cm]{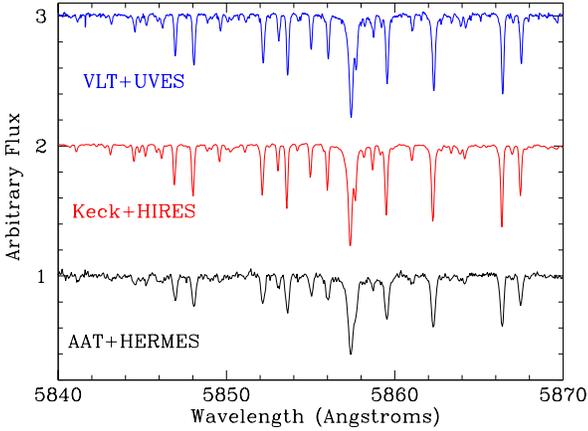}
\caption{A comparison of HERMES spectra from the Green channel in nominal resolution mode for the Hyades dwarf HIP13976 against those of the high-resolution spectrographs in 8-10m class telescopes. Note the resolution of UVES and HIRES spectra are R $\sim$60,000. The SNR is approximately 150, 250 and 80 per pixel for UVES, HIRES and HERMES spectra respectively.} 
\label{fig:comp}
\end{figure}

HERMES commissioning observations included several Hyades open cluster members that also have spectra obtained with VLT+UVES \citep{uves} and Keck+HIRES \citep{hires} available from those facilities' archives. Note the resolution and SNR of the HIRES and UVES spectra are much higher than that of the HERMES spectrum. Both the HIRES and UVES spectra are of resolution R $\sim$60,000, compared to R $\sim$28,000 for HERMES. The HIRES spectrum has a SNR $\sim$250 per pixel, the UVES spectrum has a SNR $\sim$150 per pixel and the HERMES spectrum has a SNR $\sim$80 per pixel. The effect of resolution in the observations is shown visually in Figure \ref{fig:comp} where we plot a region of the HERMES Green channel for the Hyades star HIP13976.

The GALAH data sample will provide accurate radial velocities as HERMES is housed in a clean, temperature-controlled room and mounted on vibration isolators to maintain stability. A set of 5 exposures were taken with HERMES over 6 nights for the velocity-stable star, HD 1581, which has radial velocity variations of 1.26 \ms\ over a time span of 2566 days as confirmed by HARPS spectra \citep{pepe}. For each epoch, radial velocities were determined by Fourier transform cross correlation against the first epoch of observations, making use of the IRAF fxcor packages.  

We find the measured radial velocity varies between epochs at the level of about 300\ms, which is also the currently achieved wavelength calibration accuracy. The larger uncertainty in wavelength calibration is primarily due to lack of accurately known wavelengths for the Xe lines, in the ThXe calibration lamp used in HERMES. We expect that the wavelength calibration will increase accuracy by the use of a ThAr calibration lamp together with the ThXe. The expected wavelength accuracy with the ThAr and ThXe combination, is of the order 0.002\AA. With further refined wavelength calibration or differential radial velocity determinations, HERMES will be capable of achieving radial velocity stability at the level of 100\ms.

\begin{figure}
\centering
\includegraphics[angle=270,width=8cm]{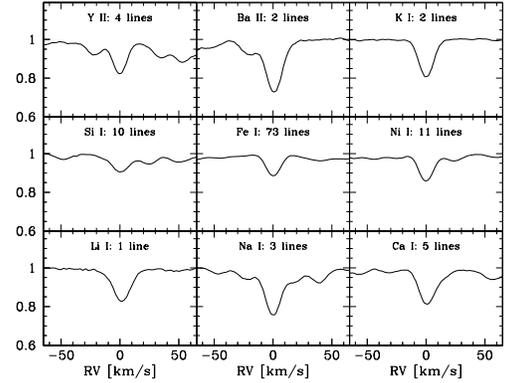}
\caption{Averaged line profiles for several representative elements in the Gaia benchmark star $\beta$ Hyi (\teff\ =5873K, \logg\ =3.98, [Fe/H] = -0.07, radial velocity = 23.16 \kms) observed with HERMES in nominal resolution mode.} 
\label{fig:benchmark}
\end{figure}

HERMES spectra contain absorption lines from about 29 elements from the major element groups and nucleosynthetic processes, including light proton-capture elements Li, C, O; $\alpha$-elements Mg, Si, Ca, Ti; odd-Z elements Na, Al, K; iron-peak elements Sc, V, Cr, Mn, Fe, Co, Ni, Cu, Zn; light and heavy slow neutron capture elements Rb, Sr, Y, Zr, Ba, La; and rapid neutron capture elements Ru, Ce, Nd and Eu.  Figure \ref{fig:benchmark} plots averages of normalized spectral tracings around lines of selected elements in radial velocity space (after shifting to zero velocity) in the Gaia benchmark star $\beta$ Hyi observed by GALAH. In each panel, the species and the number of lines considered are given at the top.

To make an initial assessment of the abundance accuracy achievable from GALAH quality spectra, equivalent widths (EWs) were measured from the HERMES solar spectra for 54 Fe I lines and 5 Fe II lines compiled by the GALAH abundance analysis group. Note that the simple analysis of the solar spectrum presented below aims only to demonstrate the quality of GALAH data. It is not representative of the GALAH pipeline analysis that will carry out a refined solar analysis and will be used to handle a large variety of spectral types. 

The measured EWs were used to derive spectroscopic parameters for the solar spectra using the 2012 version of the MOOG code \citep{moog,sobeck}. Abundances were generated for a grid of interpolated Kurucz model atmospheres based on the ATLAS9 code \citep{Castelli97} with no convective overshooting, with solar composition spanning 3500K $<$ \teff\ $<$ 8000K in steps of 50K, 0.5 $<$\logg\ $<$  5.0 in steps of 0.1 and 0.5 $<$ micro-turbulence $<$ 2.5 in steps of 0.2. The global minimum over the resulting distribution of Fe I abundance vs. excitation potential and ionisation balance, corresponding to the HERMES solar parameters were found to be \teff\ = 5750K,  \logg\ = 4.5 and micro-turbulence = 0.75, where the slope in abundance vs. excitation potential was = -0.003 dex/eV, slope in reduced width = 0.000 and log $\epsilon$ (Fe I) = log $\epsilon$ (Fe II) = 7.57 $\pm$ 0.05 dex. Note the solar abundance of Fe varies in the literature, depending on factors including the choice of atomic line data and model atmospheres, the consideration of departures from local thermodynamic equilibrium (LTE) and/or 3D hydrodynamical calculations, and the resulting solar parameters. The most commonly adopted reference solar Fe abundance based on realistic 3D photospheric models and non-LTE analysis is log $\epsilon$ (Fe I) = 7.50 $\pm$ 0.04 dex \citep{asplund09}.

\subsection{Calibration and Synergistic observations}
\label{sec:synergy}

We need to observe a number of calibration objects to ensure that we can
transform the system of the GALAH data products to the systems of other
observers. Experience from previous surveys has shown how important it is
to cover the parameter space of the stars as completely as possible. For
cross calibration with other major surveys, such as APOGEE \citep{apogee}
and Gaia-ESO \citep{gaia-eso}, we will observe a number of fields observed
by these surveys, constrained by our limiting magnitude. 

For radial velocity calibration, we plan to observe a wide selection of
stars from the Gaia radial velocity standards list. Additionally, the Red and IR channel data 
contain many intrinsically narrow telluric lines (O$_{2}$ and H$_{2}$O) that act as zero velocity 
absorption lines. Such zero-point calibration has been successfully applied in high 
accuracy radial velocity work \citep[e.g.][]{munari01}. For stellar parameter calibration, we use a grid of well measured individual stars including the Gaia benchmark stars, covering a range of temperature, gravity and metallicity, as well as a set of well observed open and globular clusters for further parameter validation. 

The CoRoT space mission \citep{corot} and the extended Kepler mission, K2 \citep{k2}, provide key opportunities for galactic archaeology with GALAH. The asteroseismology coming from these missions are very
well-matched with the spectroscopy from GALAH. From the asteroseismology of solar-like oscillations in red giants, stellar radii and masses will be measurable to a few percent accuracy \citep[e.g.][]{stello,gai}. These accurate stellar masses lead to stellar ages with precisions of better than 15-20\%. However, this is only achievable when accurate stellar temperatures and metallicities are available from spectroscopic surveys such as GALAH. Simultaneously, the seismically determined \logg\ values, which typically have uncertainties of only 0.01-0.03 dex \citep[e.g.][]{casagrande14,chaplin,Pinsonneault}, will serve as calibration for the spectral analysis of GALAH data. 

Stellar ages are notoriously difficult to determine, especially for giants that probe greater distances \citep{soderblom2010}. Having such unprecedented age information from seismology in different galactic locations gives unique insight to the formation history of the Milky Way \citep[e.g.][]{miglio2013}. With accurate ages, metallicities and velocities, these stars provide excellent opportunity to study the age-metallicity-velocity relation of the thin disc. Indeed, almost 40\% of the K2 stars to be observed so far have been proposed within the GALAH collaboration, and GALAH is on track to assemble spectra of order 40,000 K2 targets distributed across all components of the
Galaxy during K2's initial 2-year mission ending mid 2016. GALAH will observe two well-studied CoRoT fields (currently over 2000 stars observed), that are also observed by the APOGEE survey, which has already demonstrated the potential for seismic-spectroscopic collaboration in the original Kepler field
\citep{Pinsonneault}. The APOGEE-GALAH overlap will play an
important role in cross-calibration of the optically based GALAH and
infrared-derived APOGEE abundance scales.  

We highlight the complementarity of Gaia \citep{gaia} and GALAH. From
about 2017, Gaia will provide precision astrometry for the GALAH stars. At
the bright magnitudes of the GALAH sample, Gaia will be at its best, 
with parallax and proper motion uncertainties about 25 $\mu$as and
15 $\mu$as yr$^{-1}$ respectively at the end of the Gaia survey. These correspond 
to $\sim$2\% distance errors at a distance of 1 kpc and $\sim$1\kms\ errors in transverse 
velocity at 15 kpc. Combined with GALAH radial velocities (accurate to within 0.5\kms), 
the outcome will be very accurate absolute magnitudes and 3D velocities for 
the whole GALAH sample. This provides 6D phase-space coordinates, and enables the study of Galactic orbits and potential. Further, GALAH and Gaia combined provides the ability to construct accurate
colour-magnitude diagrams for all relic systems that are recovered by chemical tagging. This will be very
valuable for estimating ages and as an independent check that the stars of recovered chemically
homogeneous groups do have common ages.

\section{Scope of the GALAH survey}
\label{sec:scope}

\subsection{Survey simulations}
\label{sec:simulations}
The primary motivation for the GALAH survey is the chemical tagging
experiment described in Section \ref{sec:chemicaltagging}. Our goal is
to identify debris of disrupted clusters and dwarf galaxies. About 60\% of the GALAH stars will be dwarfs of the thin and thick disc (Table \ref{tab:sample}). The GALAH horizon for these stars is about 1 kpc. 
We assume here that all of the disrupted objects whose orbits pass through a
$\pm$1 kpc-wide annulus around the Galaxy at the solar circle are
represented within the observable horizon. Simulations
\citep{bhf04,bh2010} show that a random sample of a million stars with
V $<$ 14 will allow detection of about 20 thick disc dwarfs from each
of about 3000 star formation sites, and about 10 thin disc dwarfs from
each of about 30,000 star formation sites. Is it possible to detect
the debris of about 30,000 different star formation sites, using
chemical tagging techniques? Are there enough independent cells in
C-space to make this possible? The GALAH survey data will test these
expectations.

The simulated numbers above depend on the details of the
initial cluster mass function (ICMF) of the disrupted objects
and its mass range. Lets assume the ICMF to be a power law
specified by minimum mass x$_{\rm{min}}$, maximum mass x$_{\rm{max}}$
and power law index $\gamma$. Our expectation is that the thin disc
has a steep ICMF (large $\gamma$) and low x$_{\rm{max}}$ typical of a
quiescent star formation history and that the thick disc will have a
shallower ICMF (small $\gamma$) and large x$_{\rm{max}}$
representative of the turbulent high-pressure discs seen at high
redshift. The size of the GALAH survey is selected to probe the slope
$\gamma$ and x$_{\rm{max}}$ of the ICMF.

\begin{figure}
\centering
\includegraphics[width=9.5cm]{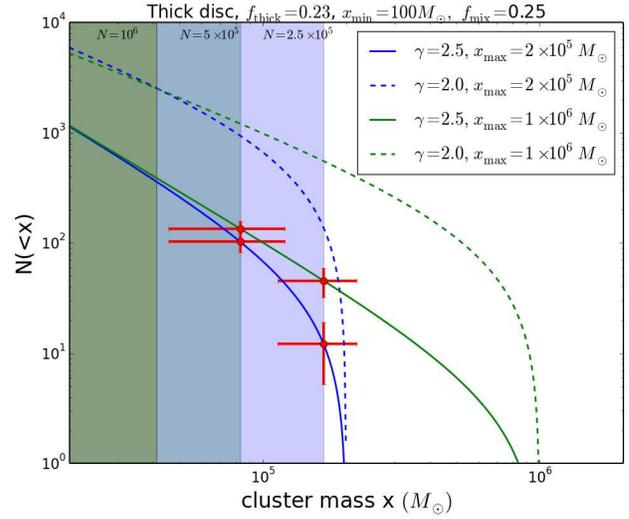}
\caption{The number of clusters recovered from a simulated GALAH
  survey as a function of initial cluster mass.  f$_{\rm pop}$ is the
  fraction of stars in a survey that belong to a given population and N is the number of survey stars. The green shaded region represent the GALAH detection limits. The dark blue and light blue shaded regions are the ranges of initial cluster mass inaccessible to surveys with 500,000 stars and 250,000 stars, respectively. The red dots show the cluster masses from which we expect to recover 20 stars (points on the left) and 40 stars (points on the right) in a million-star survey, and the number of such clusters we would expect to find if x$_{\rm{max}}$ is 2 $\times$10$^{5}$ (points on the blue curve) or 1$\times$10$^{6}$ (points on the green curve). The red error bars show 2$\sigma$ Poisson uncertainty on the number and on the size of the recovered groups.}
\label{fig:simulation}
\end{figure}

We simulate the GALAH survey based on the Galaxia tool \citep{galaxia}
and then applied a simple analytical prescription for generating
stellar clusters.
Figure \ref{fig:simulation} shows the number of unique clusters
identified in this simulated survey as a function of initial cluster
mass. Setting the minimum requirement of ten member stars for a
reliable identification of a cluster, we calculate the threshold
cluster mass (i.e. the lowest-mass cluster we expect to recover ten
stars from) as m$_{Th}$ =10 $\times$ (M$_{\rm pop}$ $\times$
f$_{\rm mix}$)/(f$_{\rm pop}$ $\times$ N), where M$_{\rm pop}$ is the total mass
of the galactic population (thin or thick disc),
N is the number of stars in the survey and
f$_{\rm pop}$ is the
fraction of stars in the survey that belong to the population.
The f$_{\rm mix}$ is the fraction of star forming mass of the population that lies
within the survey volume. If there is no mixing $f_{\rm mix}=M'_{\rm pop}/M_{\rm pop}$, 
with $M'_{\rm pop}$ being the physical mass of the population enclosed within the
survey volume. If mixing is maximal, i.e., stars born
anywhere within the Galaxy can lie in the survey volume, then 
$f_{\rm mix}=1$.
For the thick disc, M$_{\rm pop}$ =3.9 $\times$
10$^{9}$ M$_{\sun}$ and f$_{\rm pop}$=0.236, which means that m$_{Th}$=4.2
$\times$ 10$^{4}$ for N=10$^{6}$ and we assume f$_{\rm mix}$=0.25. Clusters with
initial masses below 4.2 $\times$ 10$^{4}$ are outside the detection
limits of the survey as noted by the green shaded region.  Less
efficient radial mixing brings stars from a smaller number of clusters
into our survey volume, moving all of these thresholds toward lower
mass and making cluster identification easier.

The red dots in Figure \ref{fig:simulation} show the cluster masses
from which we would expect to recover 20 stars (points on the left)
and 40 stars (points on the right) in a million-star survey, and the
number of such clusters we would expect to find if x$_{\rm{max}}$ is 2
$\times$10$^{5}$ (points on the blue curve) or 1$\times$10$^{6}$
(points on the green curve). The red error bars show 2$\sigma$ Poisson
uncertainty on the number and on the size of the recovered
groups. From this simulation we see that a smaller survey size would
mean fewer stars per formation site, from a similar number of
formation sites and severely limits the range of cluster masses over
which we can explore the ICMF.

The above calculations assume that clustering exists in chemical
abundance space and that the clusters are well separated and
observational errors are small enough ($<$ 0.1 dex) such that they can
be detected by clustering algorithms. In reality, the detectability of
clusters in C-space depends upon the dimensionality of the C-space,
the inter-cluster separation and the observational uncertainties on
abundance measurements. These questions can only be answered with a
large enough data set that has been homogeneously analysed, such as
the GALAH survey data.

\subsection{The dimensionality of the GALAH chemical space}
\label{sec:dimensions}

Within the chemical inventory that will be made available from the GALAH survey data, the effectiveness of chemical tagging is not driven simply by the number of elements studied, \citep{ramirez14}; its power derives from studying elements with independent origins, which vary relative to each other.  \cite{ting12} made a principal component analysis (PCA) of element abundances from catalogues of metal-poor stars, metal-rich stars, open clusters and also of stars in the Fornax dSph galaxy. The PCA included detailed simulation of the effects of observational errors on the apparent dimensionality of the C-space. The outcome is that the HERMES C-space is predicted to have dimensionality of 7 to 9 for all these samples, likely depending on the dominant nucleosynthetic processes in a given population.

The principal components are vectors in the C-space of the element abundances [X/Fe], and these vectors could be identifiable with nucleosynthetic processes. The principal components are eigenvectors of the correlation matrix, and are all orthogonal in the C-space: therefore the higher components are projections on hyperplanes normal to the more prominent components. The number of significant principal components is similar for metal-rich and metal-poor stars, but the actual components are different, potentially reflecting different dominant nucleosynthetic processes for each sample. The interpretation of the first principal component (the component with the largest eigenvector) is usually clear, but it is not so obvious for the others \cite[see][for a detailed discussion]{ting12}. For example, for the sample of low-metallicity stars with $-$3.5 $<$ [Fe/H] $< -$2.0, the first principal component includes all of the neutron capture elements and the $\alpha$-elements. It is probably related to core-collapse SNe producing $\alpha$-elements plus the $r$-process contribution to neutron capture elements. The second component shows an anti-correlation of $\alpha$-elements with Fe-peak and neutron capture elements, and may be related to ''normal'' core-collapse SNe which do not contribute to the $r$-process.

\cite{ting12} also compared the C-space for open clusters, which have Galactocentric radii from 6 to
20 kpc, with the C-space for metal-rich stars in the solar neighbourhood. The C-space for the clusters
has one more dimension than the stars near the Sun. We may find that the giants in the GALAH sample, which extend out to about 5 kpc from the Sun, will also have a larger C-space dimension like the clusters.
Note that the examples discussed above, the PCA components are based on samples of a few hundred stars which had not always been homogeneously analysed. The homogeneously observed and analysed GALAH sample will help to delineate the nature of the principal components more clearly, and provide insight into the nucleosynthesis processes that lie behind the chemical evolution of the Galaxy. Using PCA techniques on such a large sample of stars, it may be possible to isolate the contributions of the various nucleosynthetic processes in each Galactic component, as a function of position and velocity, or as a function of the orbital integrals.

However the mathematical principal components may not necessarily correspond directly to astrophysical processes. They may be just residuals of correlations associated with the orthogonality of the principal components. For a clearer interpretation of the structure of C-space in terms of physical galactochemical evolution models, we need different analysis techniques that are based on physical processes \citep[e.g.][]{west}. The group and cluster finding algorithms discussed in the next section are a first step in this direction.

\subsection{Group Finding}
\label{sec:groupfinding}

The end goal of the GALAH survey is to use the abundance data, complemented by phase space
information from Gaia and GALAH, and age data from Gaia, to implement the recovery of relic
systems via chemical tagging. The algorithm development of identifying relic groups from the GALAH chemical inventory is still in its infancy, but with some recent positive demonstrations. 

\cite{mitschang13} compiled a sample of open cluster data from the literature, ending with a sample of 2775 individual abundance values from 291 stars from 22 sources, all at R $>$24,000 to mimic the HERMES resolving power. Using this data \cite{mitschang13} presented a metric to quantify the chemical difference of a pair of stars and derived an empirical probability function, which describes the probability that two stars observed with a number of similar abundances are compatible with being formed in the same star formation event. However the authors note the derived function is based on the inhomogeneous data available at the time and cluster reconstruction in a chemical tagging experiment of large scale should be calibrated against confirmed dispersing clusters. 

\cite{mitschang14} applied the above mentioned methodology in a first blind chemical tagging experiment using the \cite{bensby2014study} sample of about 700 field star abundances. They reported the identification of many  chemical groups and highlighted that this metric based method is more likely to be grouping stars with a similar evolutionary history rather than representing debris of relic groups from a common parent proto-cluster. The exercise also confirmed that ages for chemically tagged groups can be derived to high accuracy via isochrone fitting to the location of the tagged stars in a colour-magnitude diagram (CMD).  This approach is a considerable improvement compared to deriving ages for single stars from their CMD location.

Other group finding algorithms under development for exploiting the GALAH chemical inventory include the EnLink density-based hierarchical group finding code \citep{sharmajohnston}, Gaussian Mixture Models to separate groups based on their location in the chemical space and Bayesian Networks as a scheme to address any missing abundance data in cases of abundance analysis issues (Ting et al. in prep).

\subsection{Chemical tagging in the inner Galactic disc}
 \label{sec:radialmigration}
 
Although young open clusters are present in the inner Galaxy, the old ($>$ 1 Gyr) surviving open
clusters lie mostly in the outer Galaxy, beyond a radius of about 8 kpc \citep{friel95}. The absence
of old open clusters in the inner Galaxy is usually attributed to the stronger disruptive influence
of the Galactic tidal field and interactions with giant molecular clouds in the inner Galaxy. This
suggests that we may expect to find the relics of many disrupted open (and globular) clusters from the GALAH giants in the inner disc. The inner Galaxy may also contain the debris of disrupted globular clusters. The O-Na anti-correlation and other light element abundance patterns are unique to and ubiquitous in Galactic globular clusters, and will help to identify the debris that comes from globular clusters. We expect about 200,000 survey giants to lie in the inner regions of the Galaxy.

Disrupted clusters will provide a strong test of the importance of radial mixing for the evolution of
the disc. Open clusters are assumed to be on near-circular orbits when they are young. In the absence of radial
mixing effects, their dispersed debris would still be on near-circular orbits and be confined to a
fairly narrow annulus around the Galaxy. On the other hand, the influence of radial mixing would
be to spread the phase-mixed azimuthally dispersed debris of individual clusters over several kpc
in radius \citep{roskar}. In this way, the radial extent of the chemically tagged debris
of disrupted clusters of various ages will give a direct measure of how important radial mixing has
actually been to the evolution of the Galaxy over cosmic time.

In this environment of disrupted clusters, we may find chemical streams delineated by the debris of
globular clusters that were in relatively eccentric orbits, or by the debris of open clusters that have
been strongly affected by radial migration. It may be possible to expose such chemical streams by
using smooth chemical evolution models \citep[e.g.][]{Chiappini} to subtract off the background abundance gradient in different elements within the framework of the Galaxia tool \citep{galaxia}.

\subsection{Origin of the Thick disc}
\label{sec:thinthick}
The origin of the thick disc is a matter of ongoing debate \citep[e.g.][]{abadi,villalobos,brook,loebman,bovy12}, with suggestions that it was accreted, heated from the thin disc by interactions with satellite galaxies, formed in situ, built through radial migration and is a monotonic thickening of the disc. From extragalactic studies, we know that thick discs are common in spiral galaxies, with varying degrees of kinematic connection to their corresponding thin discs \citep[e.g.][]{yoachim} and a potential correspondence to massive star clusters in 
redshift $\sim$2 galaxies \citep{genzel}. 

In order to understand the formation history of the thick disc component, we need to accurately measure its defining properties. The thick disc is not observationally well constrained as the spatial and kinematic overlap with the Galactic thin disc makes it difficult to distinguish the thick disc stars in small samples of data within the solar neighbourhood. Thick disc properties including its scale length, dynamical discriminators such as velocity dispersion and rotational lag as well as detailed chemical properties e.g. whether or not the thick disc exhibit any radial or vertical metallicity gradient, continue to be a major topic of discussion. However high-resolution abundance studies \cite[e.g.][]{fuhrmann,haywood, bensby2014study} show that thick disc stars are enhanced in $\alpha$-elements relative to thin disc stars of similar [Fe/H]. 

With over 20\% of the sample expected to belong to the thick disc, GALAH will provide a sample many orders of magnitude larger, extending radially and vertically over a few kilo parsecs to provide a direct estimate of the thick disc metallicity gradients as well as detailed chemical abundances, providing insights into which of the possible origin scenarios or perhaps which combination of scenarios, is the best explanation for the present-day thick disc. It will be possible to use the combination of kinematics and chemical tagging of the thick disc to evaluate which of the possible origin scenarios or perhaps which combination of scenarios, is the best explanation for the present-day thick disc.

\subsection{Ancillary Science}

\begin{figure*}
\centering
\includegraphics[width=18cm]{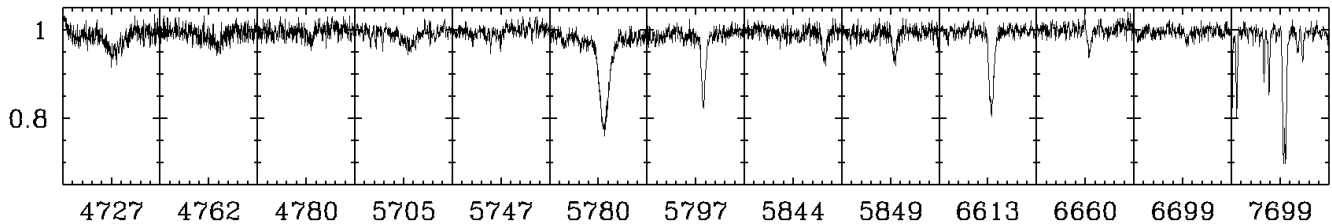}
\caption{Preliminary data on diffuse interstellar bands and on the K I interstellar atomic line at 7699\AA\ as observed by GALAH. Spectrum of TYC 4011-102-1, a hot star with strong interstellar absorptions close to the Galactic plane, is shown. Each 20\AA\ wide panel is centred on the DIB wavelength as listed in \citet{jenniskens}. Plotted wavelengths are heliocentric. Right-most panel identifies two interstellar clouds for K I at different velocities. For a majority of GALAH objects which lie away from the Galactic plane such complications are rare (but can be detected).}
\label{fig:dibs}
\end{figure*}

The GALAH sample of a million stars will have uniformly reduced and analysed high-resolution spectra, accurate radial velocities and accurate parallaxes and proper motions from Gaia. In addition to the primary motivation for Galactic archaeology, the GALAH data will be an invaluable long term resource for a huge range of stellar and interstellar science. We briefly discuss some of this GALAH science; the discussion is by no means exhaustive.

\subsubsection{Dynamic interstellar medium} 
Interstellar absorption lines give unique insight to the properties of the interstellar medium. In addition to a few atomic lines (e.g. CaII H \& K, NaI D1 \& D2, KI), the optical range also contains several hundred weak diffuse interstellar bands (DIBs) thought to be produced by organic macromolecules. \cite{kos14} found that the distribution of DIBs north and south of the Galactic plane is not equal based on RAVE spectra: the unknown DIB carrier population is not well mixed and must have been consequently perturbed fairly recently. A three-dimensional analysis of the spatial strength and radial velocities of DIBs leads to a 4-dimensional kinematic picture of the interstellar medium, pinpointing ISM bubbles arising from supernovae and other stellar explosions a few million years in the past. While this study is based on only $\sim$10000 independent sightlines from RAVE, GALAH with its higher signal-to-noise and higher spectral resolution and larger sample size will be able to recover a better spatial sampling. Figure \ref{fig:dibs} shows preliminary data on DIBs and KI line as observed by GALAH.

\subsubsection{Accurate stellar masses for binary stars}
Accurate stellar masses are scarce, where only 95 detached binary systems have both the mass and radius measured to ±3\% or better \citep{Torres10}. Such systems are important, as their components can be expected to have evolved as if they were single stars.  The Gaia mission will derive accurate astrometric orbits of the primary component with an orbital period $<$1 year \citep{2004ASPC..318..413S}.  This fixes the primary orbital period, inclination and size, but not that of the secondary orbit. The latter will not be measured by Gaia, as the instrument actually measures the motion of the photocenter, which in unequal binaries roughly coincides with the primary orbit. GALAH is in a unique position here, as it is the only survey that will obtain sufficiently high signal-to-noise and wide wavelength coverage to detect the spectrum of both binary components and measure their velocity separation. A single GALAH observation of such a system with an astrometrically determined inclination and orbital phase will yield the major axis of the relative orbit of the two components, enabling us to determine masses of both components. Such a combination of Gaia astrometry and GALAH spectroscopy will enable very accurate mass determinations for orders of magnitude more binary systems than currently known.

\subsubsection{Active stars}
Recent survey observations of open clusters (such as RAVE and Gaia-ESO) find that large fractions of open cluster stars (up to ~15\%) are active, mostly in various pre main-sequence phases of stellar evolution.   Active stars are also observed in the field, where chromospherically active stars are second only to spectroscopic binaries among the peculiar stars from RAVE \citep{matijevic2012,zerjal}. GALAH is uniquely capable of identifying activity using the spectral profiles of both H$\alpha$ and H$\beta$. Such data will help verify activity detection in marginal cases and tests the physics of the emission region via Balmer decrement. GALAH data will also be employed to investigate activity effects on spectral lines and abundance determinations.

\subsubsection{Young stellar associations and open cluster disruption}
The presence of lithium is a strong indicator of youth for pre-main sequence stars of late-G to early-M spectral types \citep{mentuch}, as lithium is fragile and is easily destroyed. Chromospheric and other types of pre-MS activity are also confined to young stellar populations. By examining the Li abundances and also emission cores in H$\alpha$ and H$\beta$ profiles, in combination with existing proper motions and the $\sim$0.5\kms\ velocities produced by the GALAH pipeline, we anticipate the discovery of hundreds of new low-mass members of young associations. This information may also enable identification of stellar ejecta from still bound or recently decomposed open clusters.

\subsubsection{Li-rich giants}
All phases of canonical stellar evolution should destroy lithium, but a small fraction of post-main-sequence stars have Li abundances higher even than primordial Big Bang Nucleosynthesis  levels. In a survey as large as GALAH, we expect to find many more Li-rich giants, leading to a more systematic understanding of their properties and the source of their lithium enrichment.

\subsubsection{Magellanic Clouds}
GALAH does not specifically target the Magellanic Clouds, but neither are they avoided. Stars with absolute magnitudes brighter than M$_{V}=-4.5$ and $-5.0$, at the distance of the Large and Small Magellanic Clouds respectively, will fall within the GALAH apparent magnitude selection. Such stars have been tentatively identified in the data taken to-date. These mostly young objects will provide information related to their star formation sites and help identify a coherent chemical tag. Such data will also allow investigations into whether these stars are "runaway" stars and explore the radial extent of the Large and Small Magellanic Clouds. There is also potential to identify luminous blue variables (supernova precursors) in the Large Magellanic Cloud.

\subsubsection{Globular clusters}
While recent work \citep[e.g.][]{milone,carretta2014} has firmly established that the stars within Galactic globular clusters can be separated into two or more distinct populations using particular elemental abundances or carefully chosen photometry, the chemical tagging approach will allow a more integrated search for population complexity within globular clusters. Also stars that have escaped from globular clusters can be identified in the field by chemical tagging \citep[e.g.][]{martell10}. Stars in globular cluster tidal tails or extra-tidal "halos" \citep[e.g.][]{marino14} will have chemistry, radial velocity and stellar parameters all in common with stars still in the cluster, allowing a study of the internal and external processes driving cluster mass loss. 

Globular clusters have long provided vital constraints on stellar evolution theory. When GALAH abundance information is combined with (i) the globular cluster multiple population phenomenon and (ii) the asteroseismic information from the Kepler K2 mission (Section \ref{sec:synergy}, K2 fields also contain globular clusters), it is clear that a new era of precision stellar astronomy will be upon us. This will make great demands of stellar theory, forcing substantial improvements in the models. Conversely, the observations will serve as very strong multidimensional constraints on theory \citep[e.g.][]{2014ApJ...781L..29B,campbell}. 

\subsubsection{Metal-poor stars}
Metal-poor halo stars provide information on star formation at early epochs \citep[e.g.][]{2013MNRAS.432L..46S}, and may even show the chemical pollution of Population III supernova, constraining the properties of the first generation of stars \citep{heger10,heger02}, e.g., whether we may find single stars or binary stars, or how fast the stars rotate, or whether there was even a significant contribution from very massive \citep{aoki14} or supermassive \citep{chen14} primordial stars. We estimate that $\sim1\%$ of GALAH stars will be halo stars (Table \ref{tab:sample}). Of these $\sim20\%$ may be metal poor with [Fe/H]$< -2.0$ \citep[see e.g.][]{2013ApJ...763...65A}. 

Whereas the lowest-metallicity halo stars \citep[e.g.][]{caffau} are likely uniquely explained by coming from a very early generation, one or a few supernovae highly diluted by primordial gas, most of the metal likely end up in stars that may be less extreme in terms of absence of metals.  Several halo stars of low [Fe/H] have a significant enhancement of CNO relative to Fe, though their very low Fe abundance \citep{keller14,frebel,christlieb} makes it hard to explain these stars as coming from much later epochs due to the increased Fe 'pollution' of the gas. The metal from Pop III stars may have ended up in stars of [Fe/H] $\sim -2$ or even above, and may be even in the bulge or thick disk rather than the halo \citep[e.g.][]{aoki14,tumlinson}. Those stars are rare and much harder to find than the spectacularly iron-deficient stars.  The large sample of stars from GALAH will increase the chances of identifying those rare stars and targeted follow-up observations can verify them as descendent of Population III stars. 

\subsubsection{Characterising exoplanet hosts}
HATSouth is a wide-field photometric survey of the southern hemisphere for transiting exoplanets \citep{2013PASP..125..154B}.  It monitors stars from multiple southern sites in 128 square degree fields with 4 minute cadence in the sloan-r band.  HATSouth photometry is optimal for stars in the same magnitude ranges as the GALAH survey (10$<$V$<$14), and therefore there will be a high degree of overlap between GALAH target stars and stars with multi-month light curves generated from the HATSouth survey.  The measured depth of an exoplanet transit provides the planet's radius as a function of the host star's radius. Stellar parameters of the host star from the GALAH pipeline will aid in determining accurate radii of transiting planets and whether they lie within the habitable zone. The GALAH abundances will provide a means by which to test if the stars identified by HATSouth as hosting a close-in transiting planet differ in abundance pattern from the more general Galactic stellar population. By cross-matching our GALAH target stars with the HATSouth light curves we will be able to determine time-variable photometric information for target stars, which will reveal pulsators, eclipsing binaries, and highly active stars in the set of GALAH targets.

\subsubsection{Extinction maps} 
Accurate spectroscopic parameters from the GALAH pipeline, in combination with available photometric information \citep{2mass,henden}, will allow a detailed investigation of the shape of the interstellar extinction curve, which is well-documented to be non-standard and non-uniform for many Galactic sightlines \citep{nataf}. GALAH data will aid in improving three-dimensional maps of the extinction in the nearby Galactic disc \citep[e.g.][]{sale, munari14}, providing a major improvement over standard integrated reddening maps.

\section{CONCLUSION}
The primary science driver for the HERMES instrument was to conduct a Galactic archaeology survey of a million stars to carry out the chemical tagging experiment. HERMES brings together a unique combination of high-resolution, multi-wavelength and multi-object capability over a large 2 degree field of view and its successful commissioning on the AAT has enabled the start of the GALAH survey.

With a data set significantly larger than any other ongoing Galactic archaeology survey and a strong focus on the disc components of the Galaxy, the GALAH dataset will be well-sampled across the high-dimensional abundance space. GALAH will be able to thoroughly map the star-formation and chemical enrichment history in a broad annulus around the disc extending in and out from the solar radius by  $\sim$1 kpc in Galactocentric radius for dwarfs and $\sim$5 kpc for clump giants. The fundamental goal of GALAH is to recover the original building blocks of disc assembly in order to generate a detailed physical picture of how the Milky Way and other spiral galaxies formed and evolved.

In addition to the overarching goal of chemical tagging, the GALAH survey science drivers span a broad range across stellar and interstellar astrophysics, which are complementary to Galactic archaeology research. The GALAH collaboration look forward to providing a data set of enormous legacy value, with superior spectral resolution, abundance precision and sample size. The coming years will be an exciting time for stellar studies of the Galaxy.

\section*{Acknowledgements}
This research has been supported in part by the Australian Research Council (ARC) funding schemes (grant numbers DP1095368, DP120101815, DP120101237, DP120104562, FS110200035 and FL110100012). AH was supported by an ARC Future Fellowship grant FT120100363. DMN was supported by the ARC grant FL110100012. SLM was supported by the ARC DECRA grant  DE140100598. 



\bibliography{References}
\label{lastpage}

\newpage
\noindent \rule{8.5cm}{1pt}

\noindent
$^1$Australian Astronomical Observatory, PO Box 915, North Ryde, NSW 1670, Australia\\
$^2$Research School of Astronomy \& Astrophysics, Australian National University, ACT 2611, Australia\\
$^3$Sydney Institute for Astronomy, School of Physics, A28, The University of Sydney, NSW 2006, Australia\\
$^4$School of Physics, University of New South Wales, Sydney, NSW 2052, Australia\\
$^5$Department of Physics and Astronomy, Macquarie University, Sydney, NSW 2109, Australia\\
$^6$Research Centre in Astronomy, Astrophysics \& Astrophotonics, Macquarie University, Sydney, NSW 2109, Australia\\
$^7$Faculty of Mathematics and Physics, University of Ljubljana, Jadranska 19, 1000 Ljubljana, Slovenia\\
$^8$Computational Engineering and Science Research Centre, University of Southern Queensland, QLD 4350, Australia\\
$^9$Max-Planck-Institut f\" ur Astronomie, D-69117 Heidelberg, Germany\\
$^{10}$Monash Centre for Astrophysics, School of Mathematical Sciences, Building 28, Monash University, VIC 3800, Australia\\
$^{11}$Institute of Astronomy, University of Cambridge, UK\\
$^{12}$INAF - Osservatorio Astronomico di Padova, vicolo dell'Osservatorio 5, 35122, Padova, Italy\\
$^{13}$International Centre for Radio Astronomy Research (ICRAR), The University of Western Australia, WA 6009, Australia\\
$^{14}$Department of Physics and Astronomy, Uppsala University, Box 516, SE-751 20 Uppsala, Sweden\\
$^{15}$INAF - Astronomical Observatory of Padova, 36012 Asiago (VI), Italy\\
$^{16}$Harvard-Smithsonian Center for Astrophysics, 60 Garden Street, Cambridge, MA 02138, USA\\
$^{17}$Joint Institute for Nuclear Astrophysics, Michigan State University, East Lansing, MI 48824-2320, USA\\
$^{18}$School of Physics and Astronomy, University of Minnesota, Minneapolis, MN 55455, USA\\
$^{19}$Department of Physics and JINA Center for the Evolution of the Elements, University of Notre Dame, Notre Dame, IN 46556, USA\\
$^{20}$INAF - Osservatorio Astrofisico di Torino, 10025, Pino Torinese, Italy\\

\end{document}